\documentclass[twocolumn,showpacs,preprintnumbers,amsmath,amssymb]{revtex4}

\usepackage{graphicx}
\usepackage{epsfig}

\usepackage{dcolumn}

\usepackage{longtable}
\usepackage{longtable}

\begin{document}


\title{Level-density parameters in superheavy nuclei}

\author{A. Rahmatinejad, A. N. Bezbakh}
\affiliation{Joint Institute for Nuclear Research, Dubna, 141980, Russia}
\author{T. M. Shneidman}
\altaffiliation[Also at ]{Kazan Federal University, Kazan 420008, Russia}
\affiliation{Joint Institute for Nuclear Research, Dubna, 141980, Russia}
\author{G. Adamian}
\affiliation{Joint Institute for Nuclear Research, Dubna, 141980, Russia}
\author{N. V. Antonenko}
\affiliation{Joint Institute for Nuclear Research, Dubna, 141980, Russia}

\author{P.~Jachimowicz}
 \affiliation{Institute of Physics,
University of Zielona G\'{o}ra, Szafrana 4a, 65516 Zielona
G\'{o}ra, Poland}

\author{M.~Kowal} \email{michal.kowal@ncbj.gov.pl}
\affiliation{National Centre for Nuclear Research, Pasteura 7, 02-093 Warsaw, Poland}
\date{\today}

\begin{abstract}
We systematically study the nuclear level densities of superheavy nuclei, including odd systems, using the single-particle energies obtained with the Woods-Saxon potential diagonalization.
Minimization over many deformation parameters for the global
minima - ground states
and the "imaginary water flow" technique on many 
deformation energy grid for the saddle points,
including nonaxial shapes has been applied. The level density parameters are calculated by fitting the obtained results with the standard Fermi gas expression. The total potential energy and shell correction dependencies of the level-density parameter are analyzed and compared at the ground state and saddle point.
These parameters are compared with the results of phenomenological expression. As shown, this expression should be modified for the saddle points, especially for small excitation energy.
The ratio of the level-density parameter at the saddle point to that at the ground state is shown to be crucial for survival probability of heavy nucleus.
\end{abstract}

\maketitle


\section{\label{sec:level1}Introduction}

Considerable progress in the
experimental synthesis of new superheavy nuclei has been achieved recently \cite{Oganessian1,Oganessian2} and
further experiments on producing heavier elements are planned with the constructed new "superheavy factory" at JINR.
The success in production of superheavy nuclei
mainly depends on how strongly a hot compound nucleus, created in complete fusion reaction, is opposed to the fission process.
Thus, the survival probability, which takes into account the competition between neutron emission and fission,
plays an important role in the formation of evaporation residues.
The relative importance of these two decay modes mainly depends on the corresponding level densities.
To estimate which of the decay processes wins this competition, one should
know the level densities at the ground states and at the fission saddle points. Moreover, the hight and
position of the saddle point are crucial for estimation of survival probability of excited compound nucleus.

To calculate level densities, one can determine all eigenvalues, with their degeneracy, of the nuclear Hamiltonian and than
count how many of them are into the energy interval of interest.
Because the total number of states exponentially increases with excitation energy
above several MeV, the problem becomes treatable only statistically.
There are a number of sophisticated combinatorial methods to do this, see for example Refs.~\cite{hilary,hilary2}
where the parity, angular momentum, pairing correlations as well as collective enhancements are
explicitly treated based on the Gogny interaction. As shown in Ref.~\cite{Blin}, there is the general and exact scheme how to calculate the particle-hole
level densities for an arbitrary single-particle Hamiltonian taking into account the Pauli exclusion principle.
The spin- and parity-dependent shell-model nuclear level densities obtained with the moment method in the proton-neutron formalism is presented in Ref.~\cite{Senkov}. Direct microscopic calculation of nuclear level densities in the shell model Monte-Carlo approach is presented in Ref.~\cite{Alhassid}.
Despite this, in practical applications we still must use a number of approximations and assumptions,
and even corrections as superfluidity effect or collective rotational and vibrational enhancement.

Ultimately, the most important in practical applications is the empirical level-density parameter $a$
which is studied in this article.
One should remember that the level density at the saddle is not the same as that at the ground state.
Indeed, the energy available for occupation of the levels at the saddle point is lowered by the difference in the deformation energy
between the saddle and ground states.
Another reason is that the single-particle levels are distributed differently due to the change of geometrical shape of compound nucleus.
It is therefore crucial to find the right saddle point that governs the fission process.
To find all saddle points on energy hypercube and than choose the proper one between them is quite a demanding task.

Based on the fact that going from low to higher energies the nuclear system reverts from a paired system to a system of noninteracting fermions, one can successfully describe it with the well-known Fermi-gas model.
In this phenomenological model, the pairing effect is taken into account with a constant parameter $\Delta$.
In the Fermi-gas model, the average value of the level-density parameter, which establishes the connection between the excitation energy and nuclear temperature, is often assumed to depend linearly on the mass number $A$ \cite{Sokolov1990}.
In real situation, the level-density parameter is energy dependent and gradually reaches an asymptotic value with increasing energies above the neutron separation energy. The phenomenological expression has been introduced in Ref.~\cite{Ignatyuk1975} to determine the energy and shell correction dependencies of the level-density parameter.

The main goal of this paper is to combine the state of art methods:
imaginary water flow technique on a many deformation space for saddles and multidimensional minimisation for ground states with the
statistical approach for calculation of level-density parameters at those extreme (saddles and minima) points.
We use the BCS model to calculate the intrinsic level densities of superheavy nuclei with $Z=112,114,116,117,118,$ and 120.
As known, this formalism is successful in description of nuclear quantities such as level density and isomeric ratio \cite{Decowski1968}.
The ratio of the level density parameter at the saddle point to that at the ground state is important to calculate the survival probability of excited heavy nucleus. This ratio will be considered with the energy dependence of neutron emission probability from
excited heavy and superheavy nuclei.

\section{\label{sec:level12} Method of Calculation}

We apply a two-step approach.
In the first step of our calculation of the level density, using the macroscopic - microscopic (MM) method,
we determine all necessary minima and saddle points.
The minima are calculated by using multi-dimensional minimization while
the saddle points by
applying Imaginary Water Flow technique (IWF) on
multidimensional energy grids.
In the second stage, we use the statistical formalism allowing to estimate the level-density parameters
at these extreme points employing the deformed single-particle spectra.

\subsection{\label{sec:lev} Ground states and saddle points}

To calculate the potential energy surfaces, the MM method is used.
 In the frame of this method the microscopic energy is calculated by applying the Strutinski shell and pairing correction method \cite{STRUT67} with
 the single-particle levels obtained after diagonalization of the deformed Woods-Saxon potential \cite{WS}.
The $n_{p}=450$ lowest proton levels and $n_{n}=550$ lowest neutron levels from the $N_{max}=19$ lowest
shells of the harmonic oscillator are taken into account in the diagonalization procedure.
The standard values of
$\hbar\omega_{0}=41/A^{1/3}$ MeV for the oscillator energy and
$\gamma=1.2 \hbar\omega_{0}$ for the Strutinski smearing parameter, and a six-order correction polynomial are used in the
calculation of the shell correction.
For the macroscopic part, we use the Yukawa plus exponential model \cite{KN}
 with the parameters specified in Ref.~\cite{WSpar}.
The deformation dependent Coulomb and surface energies are integrated by using the
64-point Gaussian quadrature.

 For nuclear ground states, based on our previous tests and results \cite{Kowal2010, Jachimowicz2017_I, Jachimowicz2014},
we confined our analysis to axially-symmetric shapes
parameterized by spherical harmonics expansion of the nuclear radius truncated at $\beta_{80}$ :
\begin{eqnarray} \label{gs}
R(\vartheta ,\varphi) = c R_0\{
1 & + & \beta_{2 0} {\rm Y}_{2 0} + \beta_{3 0} {\rm Y}_{3 0} + \beta_{4 0} {\rm Y}_{4 0} \nonumber \\
  & + & \beta_{5 0} {\rm Y}_{5 0} + \beta_{6 0} {\rm Y}_{6 0} + \beta_{7 0} {\rm Y}_{7 0} \nonumber \\
  & + & \beta_{8 0} {\rm Y}_{8 0} \},
\end{eqnarray}
 where the dependence of spherical harmonics on $\varphi$ is suppressed
 and $c$ is the volume-fixing factor depending on deformation.  
In this case, the energy is minimized over 7 degrees of freedom
 specified in (\ref{gs}), by using the conjugate gradient method. To avoid
 falling into local minima, the minimization is repeated dozens of times
 for each nucleus, with randomly selected starting deformations.
 For odd systems, the additional minimization over configurations is
 performed at every step of the gradient procedure.

Triaxial and mass-asymmetric deformations are included and the IWF method is used for finding
the saddles. This allows us to avoid errors inherent in the constrained minimization approach \cite{Myers96,Moller2000,Moller2009,Jachimowicz_sec}.
This very efficient technique in the study of fission barriers was first applied in Ref.~\cite{Mamdouh1998}.
To find saddles, the energy for each nucleus is calculated on the 5D deformation grid
and then fivefold interpolated in each dimension for the IWF search.

So, in order to find the proper first saddle point we use a five dimensional
 deformation space, with the expansion of the nuclear radius:
\begin{eqnarray} \label{sp1}
R(\vartheta ,\varphi) = c R_0\{
1 & + & \beta_{2 0} {\rm Y}_{2 0} + {\beta_{2 2} \over {\sqrt{2}}} \lbrack {\rm Y}_{2 2} + {\rm Y}_{2 -2} \rbrack  \nonumber \\
  & + & \beta_{4 0} {\rm Y}_{4 0} +  \beta_{6 0} {\rm Y}_{6 0} +
 \beta_{8 0} {\rm Y}_{8 0} \} ,
\end{eqnarray}
where the quadrupole non-axiality $\beta_{2 2}$ is explicitly included.
For each nucleus we generate the following 5D grid of deformations:
\begin{eqnarray} \label{sp1grid}
\beta_{2 0} & = & \phantom {-} 0.00 \; (0.05)  \; 0.60 \nonumber \\
\beta_{2 2} & = & \phantom {-} 0.00 \; (0.05)  \; 0.45 \nonumber \\
\beta_{4 0} & = &           -  0.20 \; (0.05)  \; 0.20           \\
\beta_{6 0} & = &           -  0.10 \; (0.05)  \; 0.10 \nonumber \\
\beta_{8 0} & = &           -  0.10 \; (0.05)  \; 0.10 \nonumber
\end{eqnarray}
of $29\:250$ points (nuclear shapes);
 the numbers in the parentheses specify the grid steps. Additionally, for
 odd - and odd - odd nuclei, at each grid point we look for low-lying
 configurations blocked by particles on levels from the 10-th below to the
 10-th above the Fermi level. Then, our primary grid (\ref{sp1grid})
 was extended by the fivefold interpolation in all directions. Finally,
 we obtained the interpolated energy grid of more than 50 million points.
 To find the first saddles on such a giant gird,
 we use the IWF method
 (see e.g. \cite{Jachimowicz2017_II, Jachimowicz2017_III}).

\subsection{\label{sec:lev1} Level-density parameters}

Based on the superconducting formalism \cite{origin1}, the constants ($G_{N}$ and $G_{Z}$) of pairing interaction for neutrons and protons are
adjusted to obtain the ground-state pairing gaps ($\Delta_{N}$ and $\Delta_{Z}$) of the MM method with the following equations:
\begin{equation} \label{eq2}
N = \sum_k \left(1-\frac{\varepsilon_{N,k}-\lambda_{N}}{E_{N,k}}\tanh\frac{\beta E_{N,k}}{2}\right),
\end{equation}
\begin{equation} \label{eq3}
\frac{2}{G_{N}}=\sum_k\frac{1}{E_{N,k}}\tanh\frac{\beta E_{N,k}}{2},
\end{equation}
for neutrons and,
\begin{equation} \label{eq4}
Z = \sum_k \left(1-\frac{\varepsilon_{Z,k}-\lambda_{Z}}{E_{Z,k}}\tanh\frac{\beta E_{Z,k}}{2}\right),
\end{equation}
\begin{equation} \label{eq5}
\frac{2}{G_{Z}}=\sum_k\frac{1}{E_{Z,k}}\tanh\frac{\beta E_{Z,k}}{2},
\end{equation}
for protons at zero temperature.
The quasiparticle energies $E_{N(Z),k}=\sqrt{(\varepsilon_{N(Z),k}-\lambda_{N(Z)})^2+\Delta_{N(Z)}^2}$ are calculated using the single-particle energies ($\varepsilon_{N(Z),k}$) of the Woods-Saxon potential.
Using the obtained values of the pairing constants, the pairing gaps and chemical potentials ($\lambda_{N(Z)}$) are determined by solving Eqs.(\ref{eq2})--(\ref{eq5}) at given temperatures ($T=1/\beta$).
Then, setting the obtained values in the following equations the excitation energies ($U=U_{Z}+U_{N}$), entropies ($S=S_{Z}+S_{N}$) and intrinsic level densities ($\rho$) are calculated:
\begin{equation} \label{eq6}
E_{Z,N}(T)=\sum_k\varepsilon_k\left(1-\frac{\varepsilon_k-\lambda_{Z,N}}{E_{k}}\tanh\frac{\beta E_{k}}{2}\right)-\frac{\Delta^{2}_{Z,N}}{G_{Z,N}},
\end{equation}
\begin{equation} \label{eq7}
U_{Z,N}(T)=E_{Z,N}(T)-E_{Z,N}(0),
\end{equation}
\begin{equation} \label{eq8}
S_{Z,N}(T)=\sum_{k}\left\{\ln[1+\exp(-\beta E_{k})]+\frac{\beta E_{k}}{1+\exp(\beta E_{k})}\right\},
\end{equation}
\begin{equation} \label{eq9}
\rho=\frac{\exp{(S)}}{(2\pi)^{\frac{3}{2}}\sqrt{D}},
\end{equation}
where $D$ is the determinant of the matrix comprised of the second derivatives of the entropy with respect to $\beta$ and $\mu=\beta\lambda$.
The calculations were repeated using the single-particle level energies obtained with the Woods-Saxon potential at the saddle point.
In the BCS calculations of the saddle point the pairing constants were taken from the ground-state results.
Above critical temperature ($T_{cr}$) the pairing gap vanishes and all thermodynamical quantities revert to those of a noninteracting Fermi system.
Generally, larger density of states close to the Fermi surface at the saddle point leads to larger pairing correlation and as a consequence larger value of the critical temperature in comparison to the ground state.
In the mass region considered,
the critical temperatures for neutrons and protons are up to $0.42$ MeV at the ground state, and $0.52$ MeV at the saddle point.
The corresponding total excitation energies are $U_{cr}\approx 5.14$ MeV at the ground state and $U_{cr}\approx 11.27$ MeV (with respect to the ground state)
at the saddle point.
Fitting the calculated values of intrinsic level density at an specified excitation energy with the back-shifted Fermi gas expression
\begin{equation} \label{eq10}
\rho_{FG}(U)=\frac{\sqrt{\pi}}{12 a^{\frac{1}{4}}(U-\Delta)^{\frac{5}{4}}}\exp({2\sqrt{a(U-\Delta)}}),
\end{equation}
one can obtain the level density parameter $a(U)$ as a function of excitation energy.
In the calculations, the energy back-shifts are taken as $\Delta=24/\sqrt{A},12/\sqrt{A}$, and 0 MeV for even-even, odd and odd-odd isotopes, respectively.
Energy and shell correction ($\delta E_{sh}$) dependencies of the level density parameter can be described with the following phenomenological expression \cite{Ignatyuk1975}
\begin{equation} \label{eq11}
a(A,U)=\tilde{a}(A)\left[1+\frac{1-\exp(-(U- \Delta)/E_{D})}{U- \Delta}\delta E_{sh}\right],
\end{equation}
where $E_{D}$ is known as the damping parameter and indicates how the shell effect in the level{sp1grid} density parameter damps with increasing $U$.
We would like to emphasise that we obtain the damping parameter and the asymptotic value $\tilde{a}$ of the level density parameter by analyzing the calculated energy dependent level-density parameter with Eq.~(\ref{eq11}) for each considered nucleus separately.
However, for practical reasons, we also want to stay at the phenomenological level. To do this one can found systematics of $\tilde{a}$ as this value smoothly depends on the mass number, see Ref.~\cite{Ignatyuk1975}
\begin{equation} \label{eq12}
\tilde{a}=\alpha A+\beta A^2.
\end{equation}
The damping parameter, in turn can be approximated as:
\begin{equation} \label{eq12b}
E_{D}=A^{1/3}/\gamma_{0}.
\end{equation}
Here, $\alpha$, $\beta$, and $\gamma_{0}$ are the parameters providing the best fit of the calculated energy dependent level density parameters in the nuclei considered.

\subsection{\label{sec:lev2} Survival probability}
Finally, the most important seems to be the use of this formalism to estimate the probability of survival of the synthesized nucleus.
The survival probability of heavy and superheavy nuclei is proportional to the ratio of neutron emission width ($\Gamma_{n}$) to fission width ($\Gamma_{f}$) \cite{Vandenbosch1973}. This ratio is calculated as:
\begin{equation} \label{eq13}
\frac{\Gamma_{n}}{\Gamma_{f}}=\frac{g A^{2/3}}{K_{0}}\frac{\int_{0}^{U-B_{n}}\varepsilon\rho_{GS}\left(U-B_{n}-\varepsilon\right)d\varepsilon}{\int_{0}^{U-B_{f}}\rho_{SP}\left(U-B_{f}-
\varepsilon\right)d\varepsilon},
\end{equation}
where $g$ is the neutron intrinsic spin degeneracy, $K_{0}\approx 10$ MeV \cite{Vandenbosch1973} and $B_{n}$ and $B_{f}$ are neutron separation energy and fission barrier height, respectively. Here, $\rho_{GS}$ and $\rho_{SP}$ are the level densities calculated at the ground state and saddle point, respectively. Based on the Fermi gas model, the following analytical expression of $\Gamma_{n}/\Gamma_{f}$ can be obtained \cite{Vandenbosch1973}
\begin{eqnarray} \label{eq14}
\frac{\Gamma_{n}}{\Gamma_{f}}&=&\frac{4 A^{2/3} a_{f} \left(U-B_{n}-\Delta_{n}\right)}{K_{0}a_{n}\left[2a_{f}^{1/2}(U-B_{f}-\Delta_{f})^{1/2}-1\right]} \nonumber \\
&\times&\exp\left[2a_{n}^{1/2}(U-B_{n}-\Delta_{n})^{1/2}-2a_{f}^{1/2}(U-B_{f}-\Delta_{f})^{1/2}\right],
\end{eqnarray}
where $\Delta_{f}$ and $\Delta_{n}$ are the back-shifts in the Fermi gas level densities at the saddle point and ground state, respectively.
Assuming only neutron emission and fission decay channels, the neutron emission probability is written as
\begin{equation} \label{eq14b}
\frac{\Gamma_{n}}{\Gamma_{tot}}=\frac{\Gamma_{n}/\Gamma_{f}}{1+\Gamma_{n}/\Gamma_{f}},
\end{equation}
and strongly affected by the ratio
\begin{equation} \label{eq15}
\frac{a_{f}}{a_{n}}=\frac{a_{SP}(A,U-B_{f})}{a_{GS}(A-1,U-B_{n})},
\end{equation}
which should be calculated and discussed for the superheavy nuclei considered.

\begin{figure}[h]
\begin{center}
\includegraphics[width=0.5\textwidth] {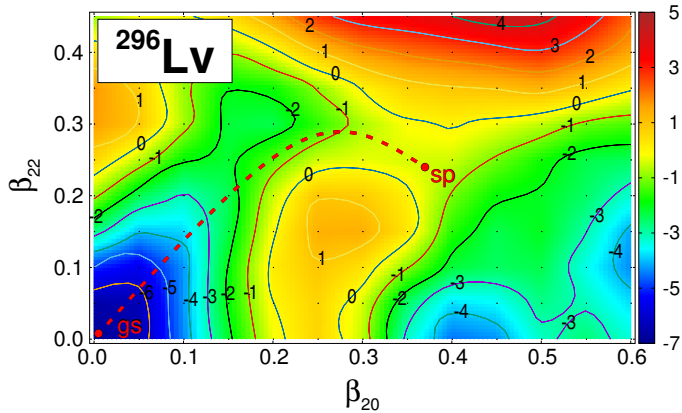}

\caption {Potential energy landscape projected into $(\beta_{20},\beta_{22})$ plane for $^{296}$Lv. Possible fission path is shown
by dashed line from the ground state (gs) to the saddle point (sp).
Energy (in MeV) is calculated relative to the macroscopic energy at the spherical shape.}  

\label{PESLv}
\end{center}
\end{figure}

\section{\label{sec:level13} Results and discussion}

\subsection{ Proton and neutron single-particle spectra}
\begin{figure}[h]
\begin{center}
\includegraphics[width=0.4\textwidth] {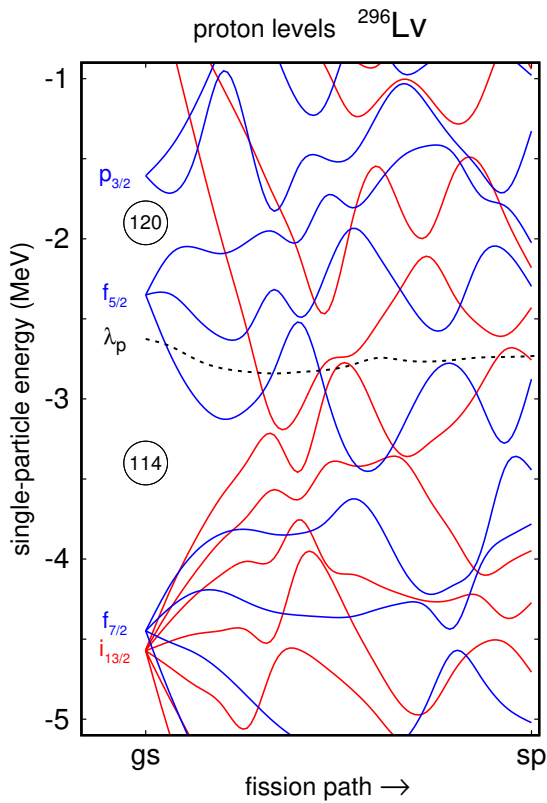}
\end{center}
\caption{Proton single-particle spectrum along the fission path for $^{296}$Lv (see Fig.~\ref{PESLv}). The Fermi level is indicated
by dotted line.}
\label{prot}
\end{figure}

\begin{figure}[h]
\begin{center}
\includegraphics[width=0.4\textwidth] {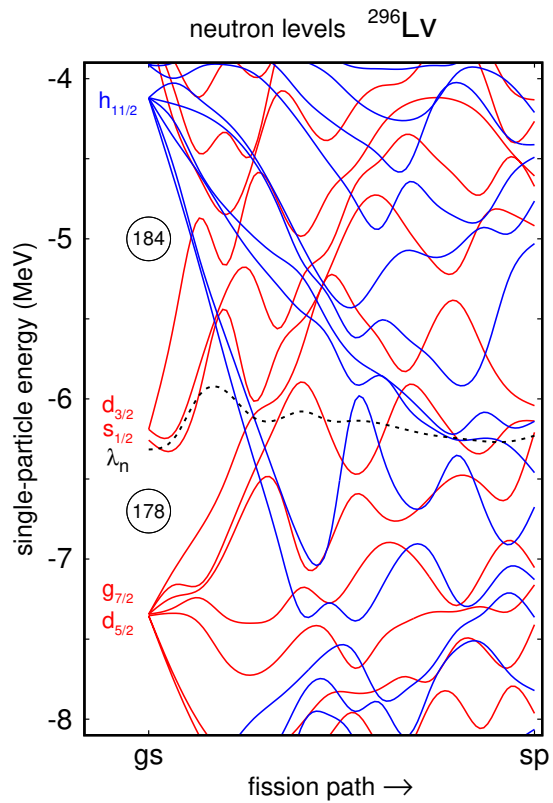}
\end{center}
\caption{The same as in Fig.~\ref{prot}, but for the neutron single-particle spectrum. }
\label{neut}
\end{figure}

As we want to compare the behavior of the level density at saddles in relation to that at the ground state,
we start with a description (giving an example) of their determination, which is crucial for further work.
As mentioned in Sect.~II, the determination of the fission barrier in multidimensional space requires
hypercube calculations and application of the IWF technique on it.
The potential energy surface for $^{296}$Lv, as an example, is shown in Fig.~\ref{PESLv} on the ($\beta_{20};\beta_{22}$) plane
obtained by minimizing energy on the five-dimensional grid (\ref{sp1grid})
with respect to $\beta_{40},\beta_{60},\beta_{80}$.
The landscape modification obtained by including quadrupole nonaxiallity deformation $\beta_{22}$
in (\ref{sp1}) is important for the picture of the first saddle points.
The strong reduction of axial barrier, by about 2 MeV, due to this effect is clearly seen on the map.
One should keep in the mind that the energy mapping in multidimensional space becomes a problem.
  A reduction of dimension via the minimization over some deformations
   often leads to an energy surface composed from disconnected patches,
  corresponding to multiple minima in the auxiliary (those minimized over) dimensions.
  This is why the real saddle point found with the IWF technique in full deformation space may be located in a slightly different
  place than the one shown in Fig.~\ref{PESLv}.
Example of fission path starting from nearly spherical ground state and ending on triaxial saddle point is shown in
Fig.~\ref{PESLv} by red-dashed line.
Along the fission path the orders of the Woods-Saxon single-particle levels are shown in Figs.~\ref{prot} and \ref{neut} for protons and neutrons, respectively.
The evolution of the Fermi level ($\lambda_{p(n)}$) is traced by black-doted lines in both cases.

Our calculations show that Eq.~(\ref{eq11}) for $a(A,U)$ gives a good agreement with the BCS calculations at the ground state, in which
the values of the shell corrections are significant.
This is fully supported by the single-particle spectra shown in Fig.~\ref{prot} for protons and in Fig.~\ref{neut} for neutrons from which
one can see the importance of shell effects.
First clearly visible is a large energy gap
for $Z = 114$ and much smaller, although still distinct for $Z = 120$ at the ground states.
This obviously contributes to the significant value of the shell effect in the ground state.
When approaching the saddle point the spectrum becomes more complicated with a lot of level crossings.
However, it is clear that the single-particle spectrum at the saddle is quite uniform and
this is why the shell correction energy is expected to be small.

Even more complicated picture of single-particle states we obtain after diagonalization of deformed Woods-Saxon potential for neutrons.
A sufficiently wide energy gap at $N = 184$ is confirmed at the ground state.
There is smaller but clear energy gap at $N = 178$. The single-particle spectrum is denser for neutrons
than that for protons.

\begin{figure}
\centering
\includegraphics[width=0.4\textwidth] {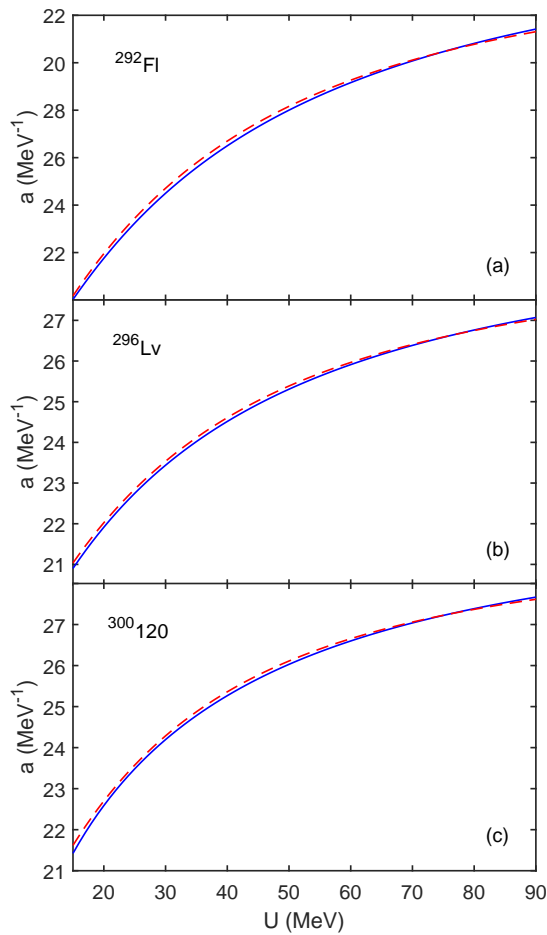}
\caption{Comparison of energy dependencies of the ground-state level-density parameters
obtained by fitting the calculated level densities (\ref{eq9}) with the Fermi-gas expression (\ref{eq10}) at the ground state (solid lines) and those obtained
 with the phenomenological expression (\ref{eq11}) (dashed lines) for nuclei $^{292}$Fl (a), $^{296}$Lv (b), and $^{300}120$ (c).}
\label{Fig1}
\end{figure}

\subsection{ Level-density parameter}

\begin{figure}[h]
\begin{center}
\includegraphics[width=0.4\textwidth] {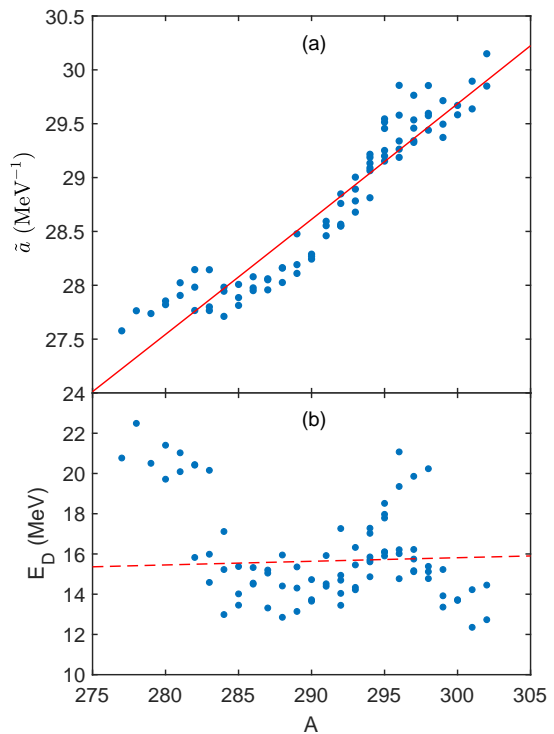}
\end{center}
\caption{(a) Mass number dependence of the asymptotic ground-state level-density parameter $\tilde a (A)$ obtained with the
phenomenological expression \eqref{eq11} (symbols). The fit of $\tilde a (A)$ with Eq.~\eqref{eq12}
is shown by solid line. (b) The corresponding damping parameters (symbols) in Eq.~(\ref{eq11}) are approximated by $E_D=A^{1/3}/0.423$ (dashed line).}
\label{Fig8}
\end{figure}

The comparisons of energy dependencies of the level-density parameter $a(A,U)$ from the fit of the calculated level densities with Eq.~(\ref{eq10}) and
those obtained from Eq.~(\ref{eq11})
are presented in Fig.~\ref{Fig1} for $^{292}$Fl,$^{296}$Lv, and $^{300}120$ nuclei.
As seen, there is a very good agreement of these dependencies at $U\geq 15$ MeV.
Analysing mass number dependence of the asymptotic level density parameters with Eq.~\eqref{eq12}, we find the coefficients
$\alpha=0.09$ MeV$^{-1}$, $\beta=2.89\times 10^{-5}$ MeV$^{-1}$ for the ground state (see Fig.~\ref{Fig8}(a)).
These values are close to those obtained in Ref.~\cite{Bezbakh2014}.
Damping parameters calculated independently for every nuclear systems at the ground states are shown in Fig.~\ref{Fig9}(b).
Despite the rather scatter nature of this parameter to use Eq.~(\ref{eq11}) one can try still to find universal value for $E_{D}$.
As found, the ground-state damping parameters are in average close to $E_{D}\approx15$ MeV.
Our results in Fig.~\ref{Fig8}(b) show that in the nuclei considered the value of $E_{D}$ can be calculated with Eq.~\eqref{eq12b} taking $\gamma_{0}=0.423$ MeV $^{-1}$. This value is close to that obtained in Ref.~\cite{Capote2009} from the analysis of neutron resonance densities and low-lying nuclear levels.

\begin{figure}
\begin{center}
\includegraphics[width=0.4\textwidth] {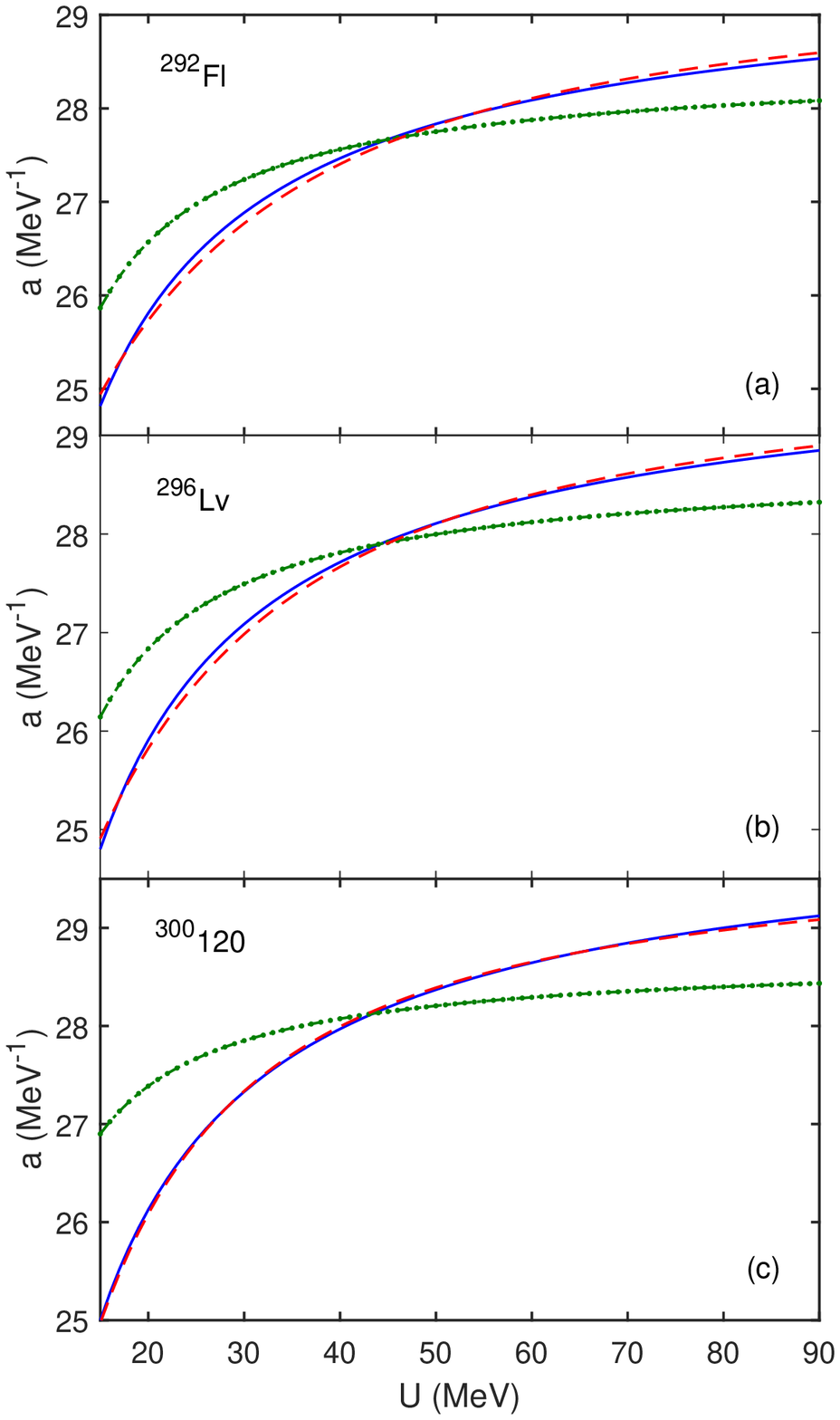}
\end{center}
\caption{The same as in Fig.~\ref{Fig1}, but for the saddle point. The calculated values of $a$ (solid lines) are compared
with those from the phenomenological expression (\ref{eq11}) with (dashed lines) and without (dash-dotted lines) replacement of
the shell correction $\delta E_{sh}$ by $\delta E_{sh}-\Delta_Z-\Delta_N$.}
\label{Fig2}
\end{figure}

\begin{figure}[h]
\begin{center}
\includegraphics[width=0.4\textwidth] {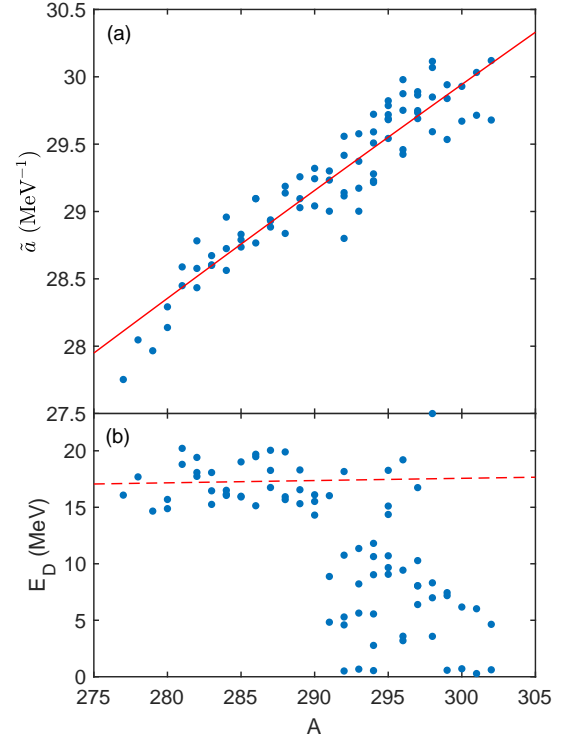}
\end{center}
\caption{The same as in Fig.~\ref{Fig8}, but for the asymptotic saddle-point level-density parameter $\tilde a (A)$ and
the saddle-point damping parameter $E_D$. Here, the damping parameters are approximated by $E_D=A^{1/3}/0.381$ (dashed line) for $A<290$.}
\label{Fig9}
\end{figure}

In Fig.~\ref{Fig2}, the energy dependencies of the saddle-point $a$ calculated for $^{292}$Fl, $^{296}$Lv, and $^{300}120$ are compared with the phenomenological model \eqref{eq11}. At the saddle point, the shell corrections are rather small or even close to zero and, thus, Eq.~(\ref{eq11}) is unsuitable to describe the calculated values of $a(A,U)$. Replacing the pure shell correction, taken as indicated just from the diagonalization of the deformed Woods-Saxon potential, in Eq.~(\ref{eq11}) by $(\delta E_{sh}-\Delta_{N}-\Delta_{Z})$, we have much better agreement with the results of direct calculations.
We obtain the following coefficients: $\alpha=0.122$ MeV$^{-1}$ and $\beta=-7.3\times 10^{-5}$ MeV$^{-1}$ in the mass dependence \eqref{eq12} of $\tilde{a}$.
It should be noted that, as in the case of the ground state, we got practically linear dependence of the parameter $\tilde{a}$
with mass number, as $\beta$ is only of the order of $\sim 10^{-5}$ in both ground and saddle points.
Comparison between the values of $\tilde{a}$ at the saddle point with the result of Eq.~\eqref{eq12} is shown in Fig.~\ref{Fig9}(a).
Damping parameters calculated independently for every nuclear systems at saddle points are shown in Fig.~\ref{Fig9}(b).
As seen in Fig.~\ref{Fig9}(b), the saddle-point damping parameters are close to $E_{D}\approx 17$ MeV for $A<290$.
Though the formal fit results in small value of $E_{D}$ at the saddle point for $A>290$, in the calculation of the survival
probability one can use larger $E_{D}$ because the shell effects at the saddle points are small in these nuclei and their damping rate
weakly influences the level density parameter in accordance with Eq.~\eqref{eq11}. Indeed, if the value of $|\delta E_{sh}|$ is small, then
the second term in the parentheses in Eq.~\eqref{eq11} is much smaller than unity.

\begin{figure}[h]
\begin{center}
\includegraphics[width=0.5\textwidth] {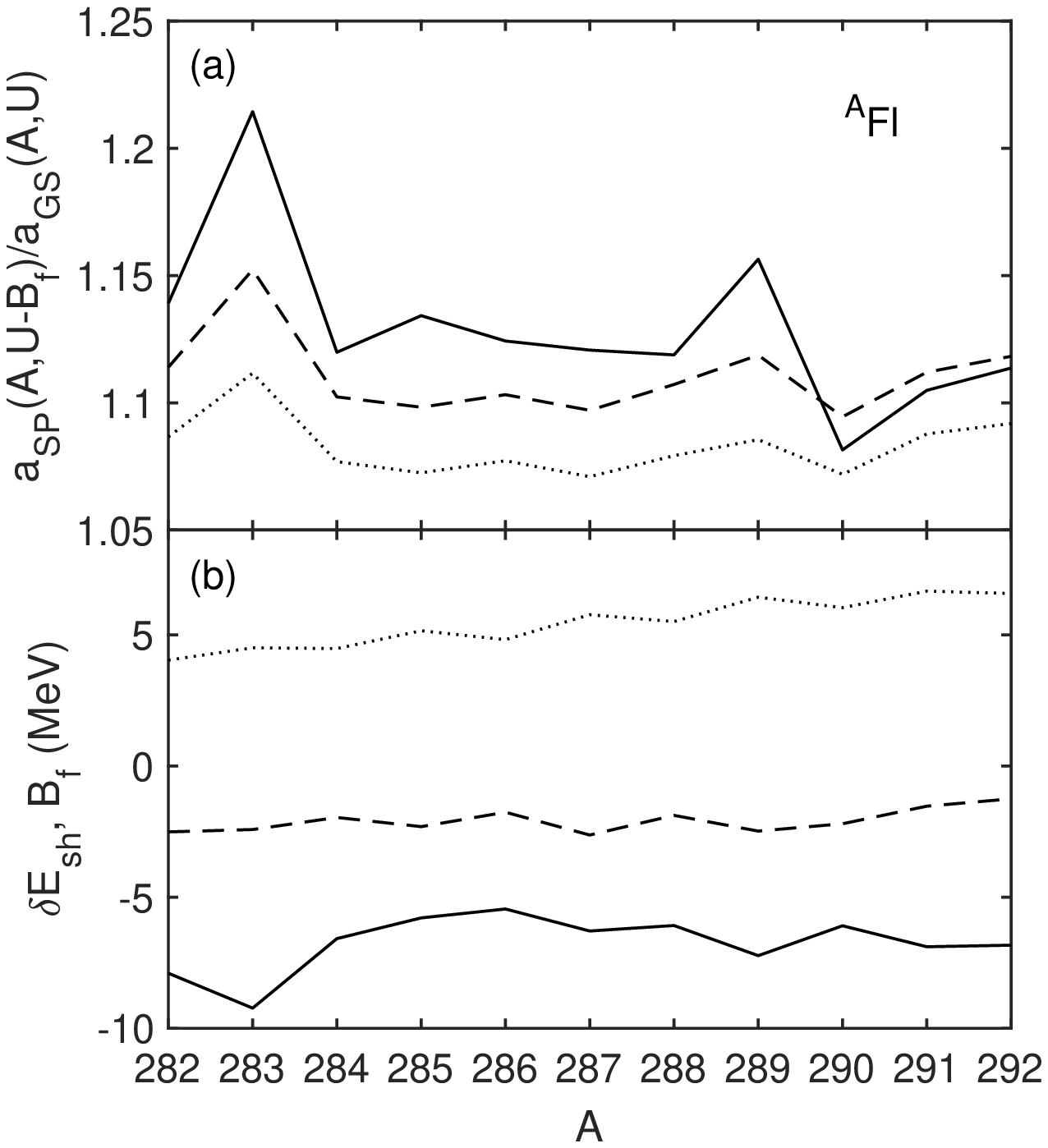}
\caption{ For $^A$Fl isotopes, (a) the ratios of the level density parameters at the saddle point and ground state
are shown at the excitation energies 20 MeV (solid line), 40 MeV (dashed line), and 60 MeV (dotted line).
At the saddle point the excitation energy is decreased by the hight $B_f$ of the fission barrier. (b) The values of
shell corrections at the ground state (solid line) and the ones at the saddle point (dashed line), and the
 heights of fission barriers (dotted line) are presented as functions of mass number $A$ at zero excitation energy.
 }
\label{114abc}
\end{center}
\end{figure}

The ratio $a_{SP}(A,U-B_{f})/a_{GS}(A,U)$ of the level-density parameter at the saddle point to that at the ground state is shown in Fig.~\ref{114abc}(a) for $Z=114$ isotopic chain at various excitation energies $U$.
The shell effects, which are evident in the $a$ ratios at $U\geq 20$ MeV, decrease at higher energies.
In Fig.~\ref{114abc}(b), the shell effects ($U=0$) are presented at the ground state (solid line) and saddle point (dashed line) for various Fl isotopes.
Thought the shell effects at the ground state are of most importance, they can not be disregarded at the saddle point.
So, the {\it topographic theorem} \cite{Swiatecki1}, that the shell effect disappear at the saddle point
and the fission barrier is just equal to the shell energy as the nucleus posses at the ground state,
is only approximately valid for superheavy nuclei.

\subsection{ Fission and neutron emission probabilities}
Because the statistical approach is used here,
the details of the fission path variability between extremes are not that important.
Only the thresholds that prohibit specific decay are required:
the separation energy with the mass of daughter nucleus
for decay via neutron emissions and the height of fission barrier.

\begin{figure}[h]
\begin{center}
\includegraphics[width=0.4\textwidth] {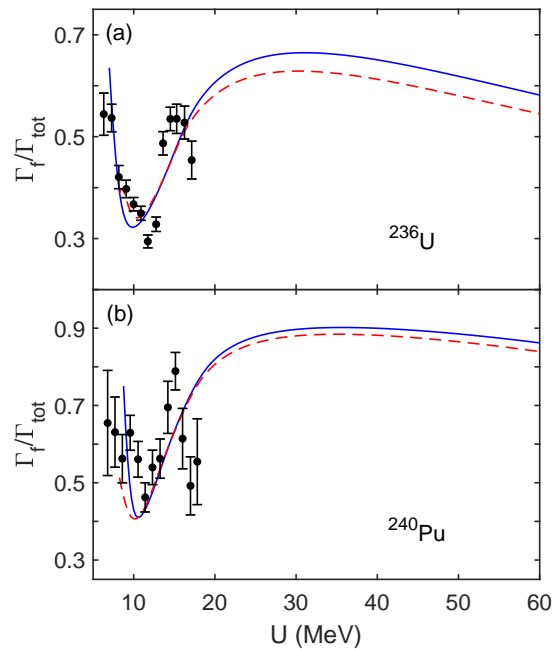}
\end{center}
\caption{The dependence of fission probability $\Gamma_{f}/\Gamma_{tot}$ on excitation energy for the fissioning nuclei $^{236}$U (a), and $^{240}$Pu (b).
 The results obtained with Eq.~\eqref{eq13} (blue solid line) and
analytical expression (\ref{eq14}) (red dashed line) are compared with the experimental data (symbols) from Ref.~\cite{Cheifetz1981}.}
\label{Fig5}
\end{figure}

To validate our calculations of the level densities, first we compare the calculated fission probabilities
$\Gamma_{f}/\Gamma_{tot}$ for $^{236}$U and $^{240}$Pu with the available experimental data.
In Fig.~\ref{Fig5}, the energy dependence of $\Gamma_{f}/\Gamma_{tot}$ calculated with Eqs.~(\ref{eq13})--(\ref{eq14b})
is shown for $^{236}$U and $^{240}$Pu together with the experimental data from Ref.~\cite{Cheifetz1981}.
In this calculation the values of $B_{n}$ are obtained from the experimental binding energies and the $B_{f}$ values are taken as the
experimental highest fission saddles, i.e., the first saddle for $^{240}$Pu \cite{Smirenkin1993}, and second saddle for $^{236}$U \cite{Capote2009}.
As seen, the
expression \eqref{eq14} gives close results to those of the numerical calculations with Eq.~\eqref{eq13}.
A good agreement with the empirical values seen in Fig.~\ref{Fig5} gives us a confidence for the
reliable  predictions of fission and neutron emission probabilities in the region of heavy nuclei.
The values of $B_{n}$ and $B_{f}$ calculated in Ref.~\cite{Jachimowicz2017_III} result in $\Gamma_{f}/\Gamma_{tot}$ values which are
less within the factor of 2, but just as dependent on energy. At $U>25$ MeV, the difference is about 10--20\%.

In the spirit of canonical Transition-State Theory (TST) \cite{Hanggi1990} the probability ratio for two selected decay types (transitions)
is proportional to the number of states available for each of them in the appropriate energy range (saturating total available energy).
What means that the excitation energy dependence of $\Gamma_{n}/\Gamma_{f}$ is strongly affected by the difference between the fission barrier height and neutron binding energy, see schematic Fig. 10.5 in \cite{Frobrich1996} or fig. 9 in \cite{KSW2005}.  Application of TST in practice one can found e.g in \cite{KSW2003,KSW2005,KSW2012,KSW2019}. As seen from Eq.~\eqref{eq13}, at $a_{f}\approx a_{n}$ an increasing $\Gamma_{n}/\Gamma_{f}$ with the excitation energy is expected at $(B_{f}+\Delta_{f})-(B_{n}+\Delta_{n})<0$, and opposite trend is expected at $(B_{f}+\Delta_{f})-(B_{n}+\Delta_{n})>0$. However, our calculations show that in the nuclei considered, the $a_{f}$ values are in average $\sim10-30\%$ larger than $a_{n}$.
\begin{figure}[h]
\begin{center}
\includegraphics[width=0.4\textwidth] {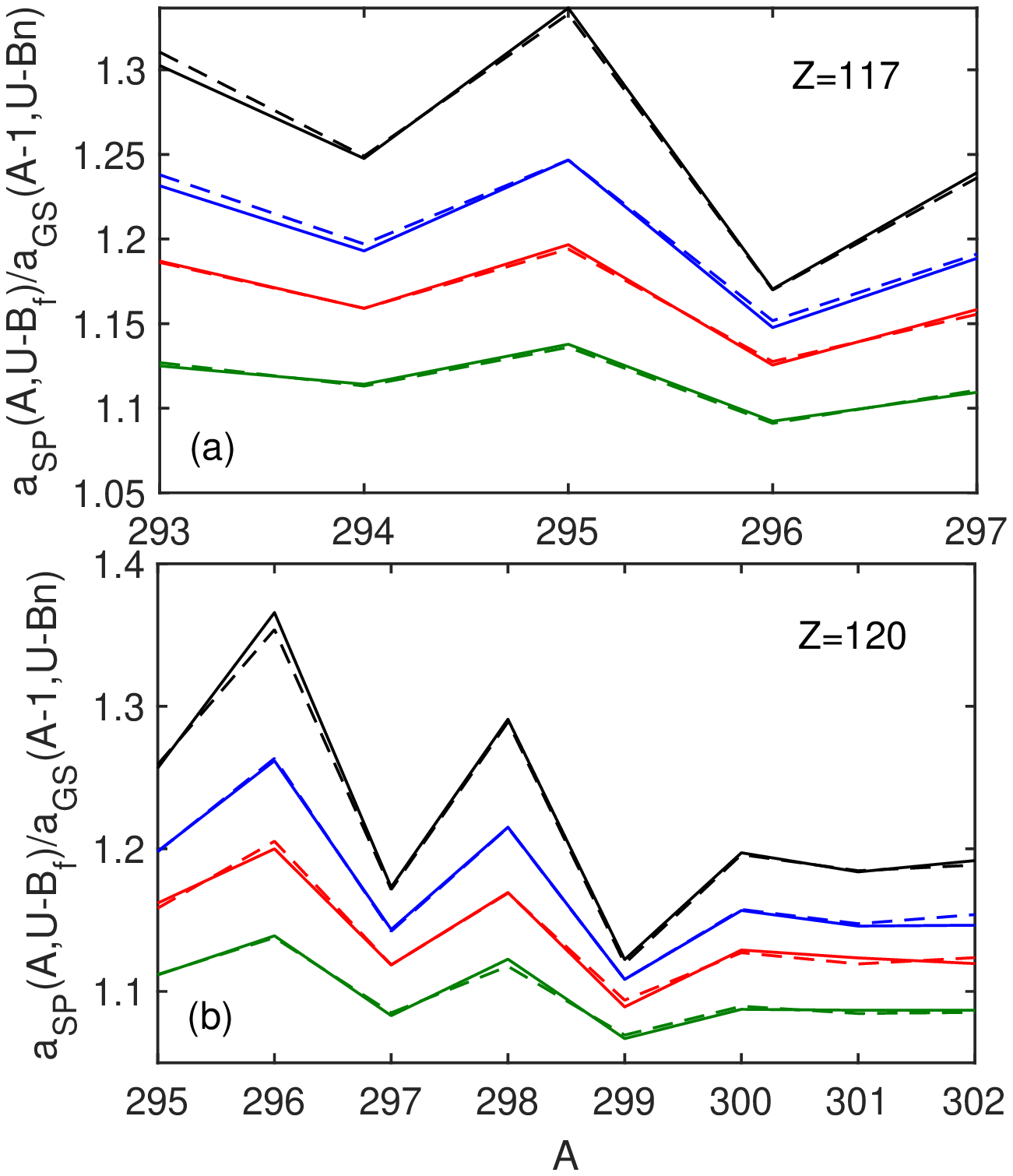}
\end{center}
\caption{For $Z=117$ (a) and $Z=120$ (b) isotopic chains, the ratios of the level density parameter of the mother nucleus at the saddle point
 to that of the daughter nucleus after neutron separation at the ground state at $U=25$ (solid black line), 35 (solid blue line), 45
 (solid red line), and 65 MeV (solid green line). The corresponding phenomenological results of Eq.~\eqref{eq11} are shown by dashed lines.
}
\label{Fig4}
\end{figure}
Because of the exponential nature of the survival probability \eqref{eq14}, the ratio of $a_{f}/a_{n}$ strongly affects the $\Gamma_{n}/\Gamma_{f}$.
The calculated ratios $a_{f}/a_{n}$ for $Z=117$ and $Z=120$ isotopic chains are shown in Fig.~\ref{Fig4}  for various excitation energies, together with the corresponding phenomenological results obtained with Eq.~\eqref{eq11}.
As seen, the shell and pairing effects decrease with excitation energy and ratios $a_{f}/a_{n}$ reach asymptotic values equal more or less 1.1.
As consequences of disappearance of quantum effects the visible strong staggering of $a_{f}/a_{n}$ weakly significantly with excitation.
One can also see that the approximate formula (Eq.~\eqref{eq11}) works very well in the whole range of excitation energy for all considered nuclei.

\begin{figure}[h]
\begin{center}
\includegraphics[width=0.4\textwidth] {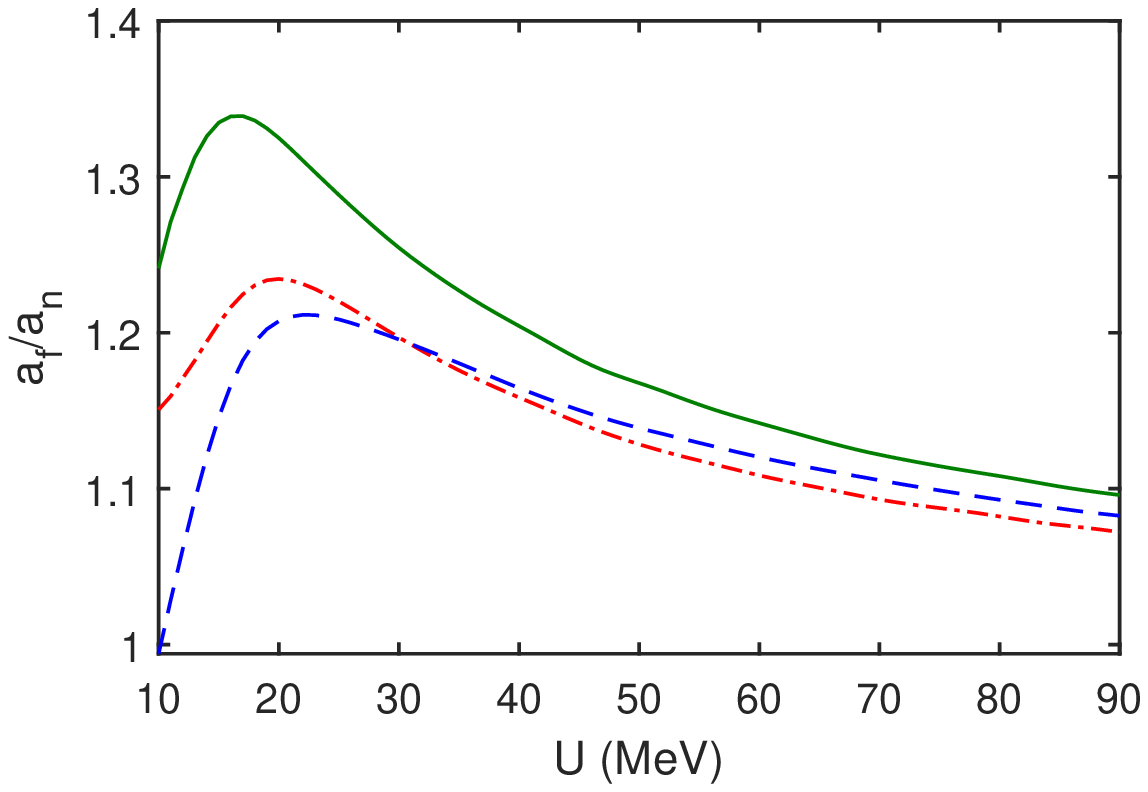}
\end{center}
\caption{The dependencies of $a_{f}/a_{n}$ \eqref{eq15} on excitation energy for $^{284}$Fl (green solid line), $^{288}$Fl (red dot-dashed line), and $^{292}$Fl (blue dashed line).}
\label{Fig4b}
\end{figure}

The dependence of $a_{f}/a_{n}$ on excitation energy for: $^{284}$Fl (green solid line); $^{288}$Fl (red dot-dashed line)
$^{292}$Fl (blue dashed line) isotopes is shown in Fig.~\ref{Fig4b}.
In all presented Fl cases $a_{f}/a_{n}$ values clearly deviate from 1, for example for $^{284}$Fl at maximum is about 1.35.
The course of variability of this ratio as a function of excitation energy is similar in all three flerovium nuclei.
At first it grows quite quickly to reach a maximum of about $20$ MeV and then slowly falls down.
Only for very high excitation energies the curves saturate to an asymptotic value of about 1.1

\begin{figure}[h]
\begin{center}
\includegraphics[width=0.4\textwidth] {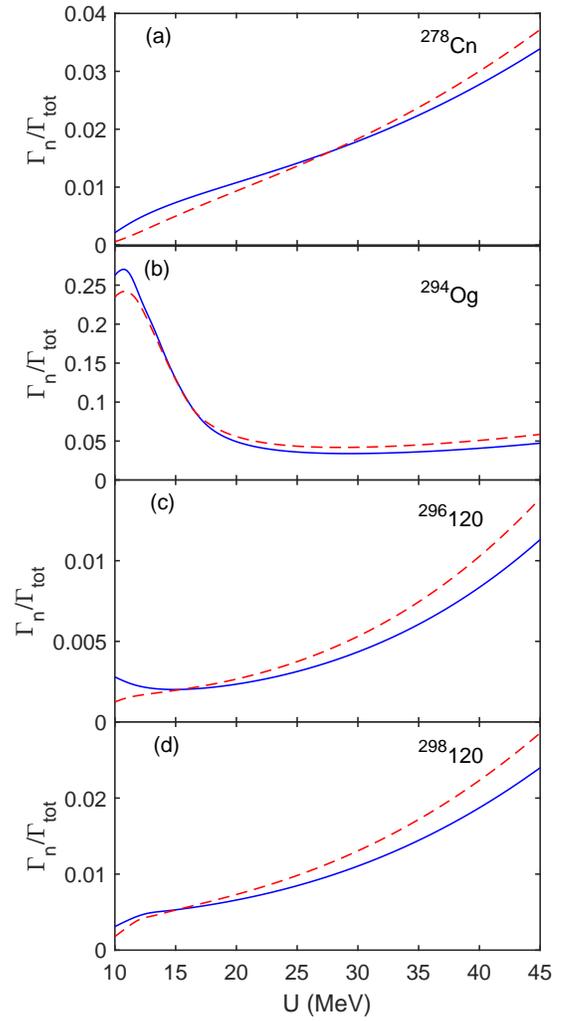}
\end{center}
\caption{The dependence of neutron emission probability on excitation energy for
$^{278}$Cn (a), $^{294}$Og (b), $^{296}$120 (c), and $^{298}$120 (d). The increasing
trend for $^{278}$Cn and $^{296,298}$120 nuclei and decreasing trend for $^{294}$Og
nucleus are in agreement with the positive or negative sign of derivative of
exponential part in \eqref{eq14}. Numerical calculations are shown by
blue solid lines and the results of expression \eqref{eq14} are depicted by red
dashed lines.}
\label{Fig7}
\end{figure}

In Fig.~\ref{Fig7}, the energy dependence of neutron emission probability is shown
for $^{278}$Cn, $^{294}$Og, and $^{296,298}$120. As follows from the expression
\eqref{eq14}, the general increase of $\Gamma_{n}/\Gamma_{tot}$ with the excitation
energy for $^{278}$Cn and $^{296,298}$120 is due to positive values of the
derivative of the exponential part
$\sqrt{a_n/(U-B_{n}-\Delta_{n})}-\sqrt{a_f/(U-B_{f}-\Delta_{f})}>0$. Similarly, the
condition
$\sqrt{a_n/(U-B_{n}-\Delta_{n})}-\sqrt{a_f/(U-B_{f}-\Delta_{f})}<0$ for $^{294}$Og
(Fig.~\ref{Fig7}(b)) nucleus leads to a decrease of $\Gamma_{n}/\Gamma_{tot}$ with increasing
excitation energy. These exact numerical results are shown in Fig.~\ref{Fig7} by solid blue lines.
Note, that depending on the superheavy system the scale of $\Gamma_{n}/\Gamma_{tot}$ is different.
The local variations seen in $\Gamma_{n}/\Gamma_{tot}$ curves at
lower energies may be caused by already discusses strong shell and pairing effects in $a_{f}/a_{n}$.
This study proves how important and even decisive is the role of energy dependent level-density parameters at the saddle point and ground state.
We have shown in
The phenomenological expression ~\eqref{eq11}
seems to be suitable for Eq.~\eqref{eq14} to describe the energy dependence of $\Gamma_{n}/\Gamma_{tot}$.
Results of these approximate calculations are shown by red dashed lines in Fig. ~\ref{Fig7}.
Only, for Z=120 both results vary noticeably for very large excitation energy values, which however, are not important to us here.

\section{\label{sec:level14}CONCLUSIONS}

Since, the nuclear level density is of special importance for the cross sections calculations and nuclear structure studies in general,
we devoted this article to this quantity.
It takes on special significance for super-heavy nuclei in which we are dealing with extremely low probabilities of their production.
The level-density parameter was evaluated here by fitting of obtained numerical and exact results to the well known Fermi gas expression.
The single-particle energies were calculated within the
microscopic-macroscopic model based on the diagonalization of the deformed single-particle Woods-Saxon potential. Yukawa-plus exponential model for macroscopic energy calculation has been used.
We carried out our analysis not only for the ground states but also for the saddle points, what's new in these kinds of calculations.
The energy and shell correction dependencies of the level-density parameter of superheavy nuclei at these extremes
were studied and compared with the well-known Ignatyuk expression.

The following conclusions can be drown:

i) We have shown that phenomenological approach based on Ignatyuk formula agrees well when one calculated energy dependence of the level-density parameter at the grounds states and strongly disagree, particularly for small excitation energies, at the saddle point configurations.
 Thus, Eq.~(\ref{eq11}) can not be safely used for the saddle at low excitation energies.

ii) Investigated, here the ratio of the level-density parameter at the saddle point to that at the ground state which is a very important factor in evaluation of the  probability of fission in comparison to the probability of neutron emissions is far from unity, especially for not to hot nuclear systems.

iii) As shown, the topographic theorem can be applied with some caution to nuclei having a substantial saddle point shell correction.

iv) We also note a substantial difference in average level-density parameters between mother nucleus at saddle point and daughter nucleus after neutron emission with expressive staggering effect visible at the same time. Having nuclear level densities at saddles and ground states one can directly/numerically
evaluate competition between neutron emission and fission. Numerically determined damping parameters have a stronger effect on the saddle compared to the ground state.

With obtained $\Gamma_{n}/\Gamma_{tot}$ one can calculate the survival probabilities of excited superheavy nuclei without involving any free parameters.
Indeed the binding energies, fission thresholds, shells corrections and finally level densities calculated on the same basis are
quite important.

\section*{ACKNOWLEDGEMENTS}

M. K. was co-financed by the National Science Centre under Contract No. UMO-2013/08/M/ST2/00257  (LEA COPIGAL).
This work was partly supported by RFBR (17-52-12015, 20-02-00176) and DFG (Le439/6-1).
We are grateful to the Polish-JINR cooperation program for its support.
T.M.S. acknowledges Russian Government Subsidy Program of the Competitive Growth of Kazan Federal University.

\end{document}